\documentstyle[12pt]{article}
\date{}

\newtheorem{Proposition}{Proposition}
\newtheorem{Theorem}{Theorem}
\newtheorem{lemma}{Lemma}
\newtheorem{Definition}{Definition}

\newcommand{\Lam}{{\Lambda}}
\newcommand{\Lamp}{{\Lambda_{\epsilon}}}
\newcommand{\lam}{{\lambda}}
\newcommand{\De}{{\Delta}}
\newcommand{\Dep}{{\Delta_{\epsilon}}}
\newcommand{\eps}{{\epsilon}}
\newcommand{\hw}{{{\hat w}_D}}
\newcommand{\w}{{w_D}}
\newcommand{\bB}{{\bar B}}
\newcommand{\Ld}{{L_D}}
\newcommand{\al}{\alpha}
\newcommand{\be}{\beta}
\newcommand{\de}{\delta}
\begin{document}
\baselineskip=14 pt

\begin{center}
{\Large\bf The Discrete AKNS-D Hierarchy}
\end{center}
\begin{center}
Xiaoning Wu\footnote{e-mail : wuxn@amss.ac.cn} \\
Institute of Applied Mathematics, \\
Academy of Mathematics and System Science,\\
Chinese Academy of Sciences,\\
 P.O.Box 2734, Beijing, China, 100080.\\
\end{center}


\begin{abstract}
In this paper, we consider the discrete AKNS-D hierarchy, find the
construction of the hierarchy, prove the bilinear identity and
give the construction of the $\tau$-functions of this hierarchy.
\end{abstract}


\section{Introduction}
The integrable system theory is an elegant achievement in
mathematics, which is developed in last century. Recently, many
authors have paid attention to consider the discrete and semi
discrete integrable system(one variable is discrete). A nice
review is given by the book edited by Bobenko and Seiler
\cite{BS}. One of the well studied topics is the q-deformed KP
hierarchy \cite{WZZ}-\cite{TCL}. In \cite{WWWY}, we studied the
q-deformation of AKNS-D hierarchy. The lax pair has been found,
the bilinear identity has been proved and $\tau$-function has been
constructed. All these results show that the q-deformed integrable
system shares similar properties with the continuous case.
However, all above results are about the q-difference operator. It
is still far away from the standard discrete system, for example,
the variable $x$ is still in ${\bf R}$ instead of in ${\bf Z}$. An
natural question is whether we can construct a discrete system
which shears the similar properties. The answer is positive.
P.Iliev considers this problem for KP hierarchy\cite{Iliev2}. In
this paper, we want to consider this problem in AKNS-D hierarchy
case.

This paper is organized as following: section II briefly reviews
the classical AKNS-D hierarchy. Section III contains the main
results of this paper. In the first part of this section, we
review some basic results of discrete calculation. In the second
part, we give the definition of discrete AKNS-D hierarchy. we also
give the Lax pair of discrete AKNS-D hierarchy. In the third part,
we define the Baker function of discrete AKNS-D hierarchy and
prove the Hirota bilinear identity. The construction of
$\tau$-function is contained by the last part. In this part, we
also consider the deformed difference operator $\Dep$ and the
relation between the discrete AKNS-D hierarchy and classical
AKNS-D hierarchy.

\section{Basic results on AKNS-D hierarchy}
We will give a brief review of AKNS-D hierarchy. Details can be
found in any standard text book.

Let $L=\partial_x-zA+U$, where $A=diag(a_1,\cdots,a_m)$ and $U$ is
a $m\times m$ matrix function of $(x,t_{k\al})$ with $u_{ii}=0,\
\forall i$. Here $k=0,1,2,\cdots,\ \al=1,\cdots,m$. The resolvent
$R$ of $L$ is defined as
\begin{eqnarray}
R&=&\sum^{\infty}_{k=0}R^{(k)}z^k,\nonumber\\
\ [L,R]&=&0.
\end{eqnarray}
All resolvent form an algebra over the field of series
$c(z)=\sum^{\infty}_{k=0}c_kz^{-k}$. This algebra has a set of
basis $\{R_{\al}\}$ which satisfies $R_{\al}^{(0)}=E_{\al}$.
$E_{\al}$ is a constant matrix and the component of $E_{\al}$ is
$\de_{i\al}\de_{j\al}$. Denote $B_{k\al}:=(z^kR_{\al})_+$, where
the subscript ``+" means taking the non-negative power terms of
$z$. The AKNS-D hierarchy is the set of equations
\begin{eqnarray}
\partial_{k\al}L=[B_{k\al},L],\label{akns}
\end{eqnarray}
where $\partial_{k\al}$ means $\frac{\partial}{\partial
t_{k\al}}$.

We can also define the dressing operator $\hat w$ of $L$ as
\begin{eqnarray}
{\hat w}(z)&=&I+\sum^{\infty}_{k=1}w_kz^{-k},\nonumber\\
L{\hat w}&=&{\hat w}\cdot(\partial_x-zA).
\end{eqnarray}
The formal Baker function $w(z)$ is
\begin{eqnarray}
w(z)={\hat
w}(z)\exp\left(\sum^{\infty}_{k=0}\sum^m_{\al=1}z^kE_{\al}t_{k\al}\right).
\end{eqnarray}
Then we have
\begin{eqnarray}
L&=&w\partial_xw^{-1},\nonumber\\
R_{\al}&=&wE_{\al}w^{-1}.
\end{eqnarray}
and the Lax pair form of AKNS-D hierarchy is
\begin{eqnarray}
L(w)&=&0,\nonumber\\
\partial_{k\al}w&=&B_{k\al}w.
\end{eqnarray}
Denote $w^*=(w^{-1})^T$, an very important property of Baker
function is the following Hirota bilinear identity.
\begin{Proposition}(Hirota bilinear identity)\\
Bilinear relation
\begin{eqnarray}
res_z[z^l(\partial_{k_1\al_1}\cdots\partial_{k_s\al_s}w)\cdot(w^*)^T]=0\nonumber
\end{eqnarray}
holds for any $l=0,1,\cdots$ and $(k_1,\al_1),\cdots,(k_s,\al_s)$.
Conversely, any two functions
\begin{eqnarray}
w&=&(I+\sum^{\infty}_{j=1}w_jz^{-j})\exp\left(\sum^{\infty}_{k=0}\sum^m_{\al=1}z^kE_{\al}t_{k\al}\right)\nonumber\\
w^*&=&(I+\sum^{\infty}_{j=1}w^*_jz^{-j})\exp\left(-\sum^{\infty}_{k=0}\sum^m_{\al=1}z^kE_{\al}t_{k\al}\right)\nonumber
\end{eqnarray}
satisfy above bilinear relation, then $w^{-1}=(w^*)^T$ and $w$ is
a baker function of an $L$ which satisfies the hierarchy equations
\ref{akns} .
\end{Proposition}

\section{The discrete AKNS-D hierarchy}
\subsection{Discrete calculations}

In this section, we want to give some basic facts about the
discrete calculations which will be needed later. We will use the
notations given by \cite{Iliev2}. In order to avoid possible
misunderstanding, we use $A(f)$ or $(Af)$ to denote an operator
$A$ acts on a function $f$ and $AB$ or $A\cdot B$ to denote the
operator multiplication.

First, we introduce two operators:
\begin{eqnarray}
\Lam(f)&:=&f(n+1),\\
\De(f)&:=&(\Lam-I)(f)=f(n+1)-f(n),
\end{eqnarray}
where $f$ is a function on ${\bf Z}$. $\Lam$ is the shift operator
and $\De$ is the difference operator. For difference operator, the
Leibnitz's Law is
\begin{eqnarray}
\De(fg)=(\Lam f)\cdot(\De g)+(\De f)\cdot g=(\De f)\cdot(\Lam
g)+f\cdot(\De g),
\end{eqnarray}
where $f$ and $g$ are functions on ${\bf Z}$.

In ref.\cite{Iliev2}, the author gives the definition of discrete
exponential function $Exp(n;t,z)$ as
\begin{eqnarray}
Exp(n;t,z):=(1+z)^n\exp\left(\sum^{\infty}_{k=1}t_kz^k\right)=\exp\left[\sum^{\infty}_{k=1}\left(t_k+n\frac{(-1)^{k-1}}{k}
\right)z^k\right]
\end{eqnarray}
Under the action of difference operator, the behavior of
$Exp(n;t,z)$ is just like the ordinary exponential function under
the action of partial differential operator, i.e.
\begin{eqnarray}
\De Exp(n;t,z)=z Exp(n;t,z).\label{exp}
\end{eqnarray}
It is easy to see that $Exp(n;t,z)$ has a formal inverse, i.e
$Exp^{-1}(n;t,z)=Exp(-n;-t,z)$. We use the word ``formal" because
$Exp(n;t,z)Exp^{-1}(-n;-t,z)=1$ but $Exp^{-1}(n;t,z)$ is
ill-defined at $z\ne-1$. What we want to emphasis is
$Exp(n;t,z)Exp^{-1}(-n;-t,z)$ is well-defined ever where. This is
an important property which we need later.

For two operators, we introduce the discrete commutator as
\begin{eqnarray}
[A,B]_D:=(\Lam A)\cdot B-B\cdot A.
\end{eqnarray}

We also need to introduce an $L^2$-inner product on the
matrix-function space over ${\bf Z}$. It takes the form
\begin{eqnarray}\label{metric}
<f,g>:=\sum^{+\infty}_{-\infty}tr(f\cdot g).
\end{eqnarray}
Under this metric, we get the dual operator of $\De$ as
\begin{eqnarray}
\De^*(f)=(\Lam^{-1}-I)(f)=f(n-1)-f(n).
\end{eqnarray}

\subsection{Discrete AKNS-D hierarchy}
Let $L_D=\De-zA+U$, where $A=diag(a_1,\cdots,a_m)$ and $U$ is a
$m\times m$ matrix function on ${\bf Z}$ with $u_{ii}=0$ for any
$i$. Like the classical AKNS-D hierarchy, we define the resolvent
$R$ for $L_D$ as
\begin{eqnarray}
&&R=\sum^{\infty}_{i=0}R_{(i)}z^{-i},\\
&&[R,L_D]_D=0.
\end{eqnarray}
Above equations can be expressed as
\begin{eqnarray}
\De R_{(i)}-[R_{(i)},U]_D+[R_{(i+1)},A]_D&=&0,\nonumber\\
\ [R_{(0)},A]_D&=&0.\label{r}
\end{eqnarray}
\begin{lemma}
All of resolvents form an algebra over the field of the formal
series $c(z)=\sum^{\infty}_{i=0}c_iz^{-i}$ and we denote it as
$\Re$.
\end{lemma}
Proof:

1) It is easy to see that $c(z)R^1+f(z)R^2$ is a resolvent of
$L_D$ if $R^1$ and $R^2$ are all resolvents.

2) If $R^1$ and $R^2$ are two resolvents of $L_D$, we have
\begin{eqnarray}
&&[R^1R^2,L_D]_D\nonumber\\
&=&(\Lam R^1)[R^2,L_D]_D+[R^1,L_D]_DR^2\nonumber\\
&=&0.
\end{eqnarray}
So we get $\Re$ is an algebra over the field of $c(z)$.$\Box$

We define the dressing operator $\hw(z)$ as
\begin{eqnarray}
\hw:=I+\sum^{\infty}_{k=1}w_kz^{-k},\label{w}
\end{eqnarray}
which satisfies
\begin{eqnarray}
L_D\hw=(\Lam\hw)\cdot(\De-zA).
\end{eqnarray}
Using the formal extension (\ref{w}), we can solve $\hw$ order by
order, so the existence of $\hw$ is obvious.

\begin{lemma}\label{multi}
$R_{\al}=\hw E_{\al}\hw^{-1}$ is a resolvent and satisfies
$R_{\al}\cdot R_{\be}=\de_{\al\be}R_{\be}$.
\end{lemma}
Proof:
\begin{eqnarray}
&&[R_{\al},L_D]_D\nonumber\\
&=&(\Lam\hw) E_{\al}(\Lam\hw)^{-1}\cdot(\De-zA+U)-(\De-zA+U)\hw
 E_{\al}\hw^{-1}\nonumber\\
&=&(\Lam\hw)
E_{\al}(\Lam\hw)^{-1}\cdot(\Lam\hw)(\De-zA)\hw^{-1}-(\Lam\hw)(\De-zA)\hw^{-1}\cdot\hw
E_{\al}\hw^{-1}\nonumber\\
&=&(\Lam\hw)[E_{\al},(\De-zA)]\hw\nonumber\\
&=&0.\\
&&R_{\al}R_{\be} \nonumber\\
&=&\hw E_{\al}\hw^{-1}\cdot\hw E_{\be}\hw^{-1}\nonumber\\
&=&\hw E_{\al}E_{\be}\hw^{-1}\nonumber\\
&=&\hw\de_{\al\be}E_{\be}\hw^{-1}\nonumber\\
&=&\de_{\al\be}R_{\be}.
\end{eqnarray}
Then we prove this lemma.$\Box$
\begin{lemma}\label{basis}
Each $R$ can be fixed by the zero order term $R_{(0)}$ and
$\{R_{\al}|\al=1,\cdots.m\}$ form a basis of $\Re$.
\end{lemma}
Proof: From the second Equation of (\ref{r}), we find $R_{(0)}$
must be diagonal and constant for $n$ (here we require all
functions will be bounded as $n\to\pm\ \infty$). Using the first
equation of (\ref{r}), we can solve each $R_{(i)}$ order by order.
The only freedom of $R_{(i)}$ is a constant diagonal part of
$R_{(i)}$ wich can be chosen as zero. So the linear independent
solutions should be those which satisfies $R_{(0)}=E_{\al}$. That
is the reason why $R_{\al}$ form a basis of $\Re$.$\Box$

We define $B_{k\al}:=(z^kR_{\al})_+$ and $\bB_{k\al}:=z^k
R_{\al}-B_{k\al}$. With these notations, we make following
definitions.
\begin{Definition}
The discrete AKNS-D hierarchy in Lax pair form is
\begin{eqnarray}
\Ld\hw&=&(\Lam\hw)\cdot(\De-zA),\nonumber\\
\partial_{k\al}\hw&=&-\bB_{k\al}\hw.
\end{eqnarray}
\end{Definition}
\begin{Definition}
The discrete AKNS-D hierarchy is defined as
\begin{eqnarray}
\partial_{k\al}\Ld=[B_{k\al},\Ld]_D.\label{dakns}
\end{eqnarray}
\end{Definition}
The relation between above two definitions of discrete AKNS-D
hierarchy is expressed in following theorem.
\begin{Theorem}
Definition 1 and definition 2 are equivalent.
\end{Theorem}
Proof:

1) If we have $\Ld$ and $\hw$ satisfy definition 1,
\begin{eqnarray}
\partial_{k\al}\Ld
&=&\partial_{k\al}[(\Lam\hw)(\De-zA)\hw^{-1}]\nonumber\\
&=&-(\Lam
\bB_{k\al})(\Lam\hw)(\De-zA)\hw^{-1}+(\Lam\hw)(\De-zA)\partial_{k\al}\hw^{-1}\nonumber\\
&=&-(\Lam
\bB_{k\al})(\Lam\hw)(\De-zA)\hw^{-1}-(\Lam\hw)(\De-zA)\hw^{-1}(\partial_{k\al}\hw)\hw^{-1}\nonumber\\
&=&-(\Lam
\bB_{k\al})(\Lam\hw)(\De-zA)\hw^{-1}+(\Lam\hw)(\De-zA)\hw^{-1}\bB_{k\al}\nonumber\\
&=&-[\bB_{k\al},\Ld]_D\nonumber\\
&=&[B_{k\al},\Ld]_D.
\end{eqnarray}

2) Conversely, if we have $\Ld$ which satisfies definition 2.
Based on the definition (\ref{w}), we get the dressing operator
$\hw$. Further more, we can also construct the resolvent
$R_{\al}$. Because $\Ld$ satisfies definition 2 and $R_{\be}$ is
resolvent of $\Ld$, we have
\begin{eqnarray}
&&[\partial_{k\al}R_{\be},\Ld]_D\nonumber\\
&=&\partial_{k\al}[R_{\be},\Ld]_D-[R_{\be},\partial_{k\al}\Ld]_D\nonumber\\
&=&-[R_{\be},[B_{k\al},\Ld]_D]_D\nonumber\\
&=&-(\Lam R_{\be})(\Lam B_{k\al})\Ld+(\Lam B_{k\al})(\Lam
R_{\be})\Ld+\Ld R_{\be} B_{k\al}-\Ld B_{k\al}R_{\be}\nonumber\\
&=&[[B_{k\al},R_{\be}],\Ld]_D,
\end{eqnarray}
i.e. $\partial_{k\al}R_{\be}-[B_{k\al},R_{\be}]$ is a resolvent of
$\Ld$. Keep lemma \ref{basis} in mind, we know
\begin{eqnarray}
\partial_{k\al}R_{\be}-[B_{k\al},R_{\be}]=\sum_{\gamma}c_{\gamma}(z)R_{\gamma}
\end{eqnarray}
Choosing any $R_{\eta}$, $\eta\ne\be$, lemma \ref{multi} tells us
\begin{eqnarray}
&&R_{\eta}(\partial_{k\al}R_{\be}-[B_{k\al},R_{\be}])\nonumber\\
&=&R_{\eta}\partial_{k\al}R_{\be}-R_{\eta}B_{k\al}R_{\be}\nonumber\\
&=&-(\partial_{k\al}R_{\eta})R_{\be}+[B_{k\al},R_{\eta}]R_{\be}\nonumber\\
&=&(-\sum_{\gamma}{\tilde c}_{\gamma}(z)R_{\gamma})R_{\be}\nonumber\\
&=&-{\tilde c}_{\be}(z)R_{\be}\nonumber\\
&=&c_{\eta}(z)R_{\eta}.
\end{eqnarray}
Because we have known that $R_{\eta}$ and $R_{\be}$ are linear
independent, we have ${\tilde c}_{\be}(z)=c_{\eta}(z)=0$, so we
have $\partial_{k\al}R_{\be}-[B_{k\al},R_{\be}]=c(z)R_{\be}$.
Repeat above method, we have
\begin{eqnarray}
&&(\partial_{k\al}R_{\be}-[B_{k\al},R_{\be}])R_{\be}\nonumber\\
&=&c(z)R_{\be}\nonumber\\
&=&\partial_{k\al}R_{\be}-[B_{k\al},R_{\be}]-(\partial_{k\al}R_{\be}-[B_{k\al},R_{\be}])R_{\be}\nonumber\\
&=&0,
\end{eqnarray}
that tells us that $\partial_{k\al}R_{\be}-[B_{k\al},R_{\eta}]=0$.
\begin{eqnarray}
&&(\partial_{k\al}z^lR_{\be}-\partial_{l\be}z^kR_{\al})_+\nonumber\\
&=&[B_{k\al},z^lR_{\be}]_+-[B_{l\be},z^kR_{\al}]_+\nonumber\\
&=&[B_{k\al},B_{l\be}]_++[B_{k\al},\bB_{l\be}]_++[B_{k\al},B_{l\be}]_++[\bB_{k\al},B_{l\be}]_+\nonumber\\
&=&[B_{k\al},B_{l\be}]+[B_{k\al},z^lR_{\be}]_++[\bB_{k\al},B_{l\be}]_+\nonumber\\
&=&[B_{k\al},B_{l\be}]+[B_{k\al},z^lR_{\be}]_++[\bB_{k\al},z^lR_{\be}]_+-[\bB_{k\al},\bB_{l\be}]_+\nonumber\\
&=&[B_{k\al},B_{l\be}],
\end{eqnarray}
so we have
$\partial_{k\al}B_{l\be}-\partial_{l\be}B_{k\al}=[B_{k\al},B_{l\be}]$.
Like the classical case, we just extend the operator
$\partial_{k\al}$ on $\hw$ by requiring
$\partial_{k\al}\hw=-\bB_{k\al}\hw$. Above result insures the
commutation relation $[\partial_{k\al},\partial_{l\be}]=0$, then
we prove this theorem. $\Box$

\subsection{Baker function and bilinear identity of discrete
AKNS-D hierarchy}

We define the Baker function of discrete AKNS-D hierarchy as
\begin{eqnarray}
\w:=\hw\cdot
g(n;t,z)=\hw\cdot(1+zA)^n\exp\left(\sum^{\infty}_{k=0}\sum^n_{\al=1}z^kE_{\al}t_{k\al}\right).
\end{eqnarray}
Using Eq.(\ref{exp}), it is easy to see that
\begin{eqnarray}
g(n;t,z)&=&\exp\left(\sum^{\infty}_{k=0}\sum^n_{\al=1}t'_{k\al}E_{\al}z^k\right),
\end{eqnarray}
where $t'_{k\al}=t_{k\al}+n\frac{(-1)^{k-1}}{k}(a_{\al})^k$. With
this definition, we have following result,
\begin{lemma}
\begin{eqnarray}
\Ld(\w)&=&0,\nonumber\\
\partial_{k\al}(\w)&=&B_{k\al}\w.\label{baker}
\end{eqnarray}
\end{lemma}
Proof:
\begin{eqnarray}
\Ld(\w)&=&(\Lam\hw)(\De-zA)\hw^{-1}\cdot\hw g(n;t,z)\nonumber\\
&=&(\Lam\hw)(\De-zA)g(n;t,z)\nonumber\\
&=&0.
\end{eqnarray}
\begin{eqnarray}
\partial_{k\al}(\w)&=&\partial_{k\al}(\hw g(n;t,z))\nonumber\\
&=&-\bB_{k\al}\hw g(n;t,z)+\hw z^kE_{\al}g(n;t,z)\nonumber\\
&=&-\bB_{k\al}\w+z^kR_{\al}\w\nonumber\\
&=&B_{k\al}\w.\qquad\qquad\qquad\qquad\Box
\end{eqnarray}
Denote $\w^*=(\w^{-1})^T$, we get
\begin{eqnarray}
\Ld^*(\w^*)=0,
\end{eqnarray}
where $\Ld^*$ is the dual operator of $\Ld$ under the $L^2$-inner
product (\ref{metric}). With the definition of Baker function, we
can prove the Hirota bilinear identity of discrete AKNS-D
hierarchy.
\begin{Theorem}(Bilinear identity)\label{bil}\\
I) If $\w$ is a solution of Eq.(\ref{baker}), it satisfies
following identity,
$$res_z(z^l(\De^m\partial^{[\lam]}_{k\al}\w)\cdot\w^{-1})=0,$$
for any $l=0,1,\cdots$, $m=0,1$ and $\forall\ [\lam]$.( Here
$\partial^{[\lam]}_{k\al}=\partial_{k_1\al_1}\partial_{k_2\al_2}\cdots\partial_{k_s\al_s}$,
and $(k_1,\al_1),\cdots,(k_s,\al_s)$ are set of index)\\
II) Conversely, if we have two functions $\w$ and $\w^*$
\begin{eqnarray}
\w&=&(1+\sum^{\infty}_{k=1}w_kz^{-k})g(n;t,z),\nonumber\\
\w^*&=&(1+\sum^{\infty}_{k=1}w^*_kz^{-k})g(-n;-t,z),\nonumber
\end{eqnarray}
which satisfy
\begin{eqnarray}
res_z\left\{z^l(\De^m\partial^{[\lam]}_{k\al}\w)\cdot(\w^*)^T\right\}=0,\nonumber
\end{eqnarray}
for any $l=0,1,2,\cdots$, $m=0,1$ and $\forall\ [\lam]$, then it implies\\
1. $\w^{-1}=(\w^*)^T$,\\
2. $\w$ is a solution of Eq.(\ref{baker}).
\end{Theorem}
Proof:

1) If $\w$ is a solution of Eq.(\ref{baker}),
$\partial^{[\lam]}_{k\al}\w=f(B_{k\al})\cdot\w$, where
$f(B_{k\al})$ is a differential polynomial of $\{B_{k\al}\}$ and
$(f(B_{k\al}))_+=f(B_{k\al})$. The bilinear identity is
\begin{eqnarray}
&&res_z[z^l(\De^m\partial^{[\lam]}_{k\al}\w)\cdot\w^{-1}]\nonumber\\
&=&res_z\left\{z^l[\De^mf(B_{k\al})\w]\cdot\w^{-1}\right\}\nonumber\\
&=&res_z\left\{z^l[\De^mf(B_{k\al})]\w\cdot\w^{-1}\right\}+res_z\left\{z^l[\Lam^mf(B_{k\al})][\De^m\w]\cdot\w^{-1}\right\}\nonumber\\
&=&res_z\left\{z^l[\De^mf(B_{k\al})]\right\}+res_z\left\{z^l[\Lam^mf(B_{k\al})](zA-U)^m\right\}\nonumber\\
&=&0.
\end{eqnarray}

2) If $\w$ and $\w^*$ satisfy the bilinear identity, we take $m=0$
and $[\lam]=0$, so we have $\w^{-1}=(\w^*)^T$ because the bilinear
identity holds for any $l\ge 0$.

Consider following equation,
\begin{eqnarray}
&&\partial_{k\al}\w-B_{k\al}\w\nonumber\\
&=&\partial_{k\al}(\hw\ g(n;t,z))-B_{k\al}\hw\ g(n;t,z)\nonumber\\
&=&(\partial_{k\al}\hw)g(n;t,z)+\hw\
z^kE_{\al}g(n;t,z)-B_{k\al}\hw\ g(n;t,z)\nonumber\\
&=&(\partial_{k\al}\hw)g(n;t,z)+z^kR_{\al}\hw\
g(n;t,z)-B_{k\al}\hw\ g(n;t,z)\nonumber\\
&=&(\partial_{k\al}\hw+\bB_{k\al}\hw)g(n;t,z).
\end{eqnarray}
The bilinear identity tells us
\begin{eqnarray}
res_z[z^l(\partial_{k\al}\w-B_{k\al}\w)\cdot(\w^*)^T]=0.\label{1}
\end{eqnarray}
Using above result, we re-write this equation as
\begin{eqnarray}
&&res_z[z^l(\partial_{k\al}\w-B_{k\al}\w)\cdot(\w^*)^T]\nonumber\\
&=&res_z[z^l(\partial_{k\al}\hw+\bB_{k\al}\hw)g(n;t,z)\cdot(\hw^*g(-n;-t,z))^T]\nonumber\\
&=&res_z[z^l(\partial_{k\al}\hw+\bB_{k\al}\hw)\cdot(\hw^*)^T]\nonumber\\
&=&0.\label{2}
\end{eqnarray}
The zero result of Eq.(\ref{1}) tells us that the terms in bracket
contain no negative power of $z$, but we can also see that there
is no positive power term in the bracket of last line of
Eq.(\ref{2}) and the zero order term of it is $0$, then we get
\begin{eqnarray}
\partial_{k\al}\hw+\bB_{k\al}\hw=0.
\end{eqnarray}
Introduce
$\Ld=(\Lam\w)\cdot\De\w^{-1}=\De+(\Lam\w)\cdot(\De\w^{-1})=\De-(\De\w)\cdot\w^{-1}$,
the formal extension of $\w$ tells us that the highest power term
in $\Ld$ is $zA$ and bilinear identity implies
$(\De\w)\cdot\w^{-1}$ contains no negative power term. The zero
order term gives $-U$, then we finish the proof.$\Box$

\subsection{$\tau$-function of discrete AKNS-D hierarchy}
In classical AKNS-D hierarchy, we can introduce $\tau$-function to
construct the Baker function $w$\cite{dic1,dic2,dic3}. The
relation between $\tau$-function and Baker function for classical
AKNS-D hierarchy is
\begin{eqnarray}
(w)_{\al\be}&=&z^{-1}\frac{\tau_{\al\be}(\cdots,t_{k\gamma}-\frac{1}{k}z^{-k},\cdots)}{\tau_D(t)},\nonumber\\
(w)_{\al\al}&=&\frac{\tau(\cdots,t_{k\gamma}-\frac{1}{k}z^{-k},\cdots)}{\tau(t)}.\label{cb}
\end{eqnarray}
Define the discrete shift of $t_{k\al}$ as
$t'_{k\al}=t_{k\al}+n\frac{(-1)^{k-1}}{k}\ (a_{\al})^k$. If $\tau$
is a $\tau$-function of classical AKNS-D hierarchy, we denote
$\tau_D(n;t)$ as $\tau_D(n;t)=\tau(t')$ and call $\tau_D$ as the
$\tau$-function of discrete AKNS-D hierarchy. Like the classical
case, we also use relation (\ref{cb}) to define $\w$, i.e.
\begin{eqnarray}
(\w)_{\al\be}&=&z^{-1}\frac{(\tau_D)_{\al\be}(\cdots,t_{k\gamma}-\frac{1}{k}z^{-k},\cdots)}{\tau_D(t)},\nonumber\\
(\w)_{\al\al}&=&\frac{\tau_D(\cdots,t_{k\gamma}-\frac{1}{k}z^{-k},\cdots)}{\tau_D(t)}.\label{db}
\end{eqnarray}
then we have following theorem.
\begin{Theorem}
The function $\w$ defined by Eq.(\ref{db}) is a baker function of
discrete AKNS-D hierarchy (\ref{dakns}).
\end{Theorem}
Proof: Because of the definition in (\ref{db}), it is easy to see
that $\hw(t)={\hat w}(t')$. We have know that the classical
$\tau$-function satisfies bilinear identity. What we need to show
is above $\hw$ satisfies bilinear identity in Theorem \ref{bil}.

For $m=0$, because $\frac{\partial}{\partial
t_{k\al}}=\frac{\partial}{\partial t'_{k\al}}$, we have
\begin{eqnarray}
&&res_z[z^l(\partial^{[\lam]}_{k\al}\w)\cdot w^{-1}]\nonumber\\
&=&res_z[z^l(\partial^{[\lam]}_{k\al}w(t'))\cdot w^{-1}(t')]\nonumber\\
&=&res_z[z^l(\partial'^{[\lam]}_{k\al}w(t'))\cdot w^{-1}(t')]\nonumber\\
&=&0.
\end{eqnarray}

For $m=1$,
\begin{eqnarray}
&&res_z[z^l(\De\partial^{[\lam]}_{k\al}\w)\cdot\w^{-1}]\nonumber\\
&=&res_z\left\{z^l[\partial^{[\lam]}_{k\al}w(t_{s\be}+(n+1)\frac{(-1)^{s-1}}{s}\
(a_{\be})^s)]\cdot w^{-1}(t_{s\be}+n\frac{(-1)^{s-1}}{s}\
(a_{\be})^s)\right\}\nonumber\\
&&
-res_z\left\{z^l[\partial^{[\lam]}_{k\al}w(t_{s\be}+n\frac{(-1)^{s-1}}{s}\
(a_{\be})^s)]\cdot w^{-1}(t_{s\be}+n\frac{(-1)^{s-1}}{s}\
(a_{\be})^s)\right\}.
\end{eqnarray}
The second term of above equation is zero because of the bilinear
identity of classical AKNS-D hierarchy. Taking the formal Taylor
extension of function $w(t_{s\be}+(n+1)\frac{(-1)^{s-1}}{s}
(a_{\be})^s)$, we get
\begin{eqnarray}
&&res_z\left\{z^l[\partial^{[\lam]}_{k\al}w(t_{s\be}+(n+1)\frac{(-1)^{s-1}}{s}\
(a_{\be})^s)]\cdot w^{-1}(t_{s\be}+n\frac{(-1)^{s-1}}{s}\
(a_{\be})^s)\right\}\nonumber\\
&=&\sum_{j,\gamma,\eta}c([\eta])\left(\frac{(-1)^{j-1}}{j}(a_{\gamma})^j\right)^{[\eta]}res_z\left\{z^l[\partial^{[\lam]}_{k\al}
\partial^{[\eta]}_{j\gamma}w(t'_{s\be})]\cdot w^{-1}(t'_{s\be})\right\}\nonumber\\
&=&0,
\end{eqnarray}
where $c([\eta])$ are Taylor extension coefficients. Above zero
result comes from classical bilinear identity because all terms in
the summation are zero. This implies the $\tau$-function got from
the shifting method is really a $\tau$-function of discrete AKNS-D
hierarchy. $\Box$

Finally, we also want to consider the relation between classical
AKNS-D hierarchy and discrete AKNS-D hierarchy. In order to do
that, we have to introduce a deformed difference operator $\Dep$
and shift operator $\Lamp$ as
\begin{eqnarray}
\Dep(f_{\eps})(n)&:=&\frac{f(x+n\eps+\eps)-f(x+n\eps)}{\eps},\\
\Lamp(f_{\eps})(n)&:=&f(x+n\eps+\eps).
\end{eqnarray}
All functions we used should be deformed as $f(n)\to
f_{\eps}(n)=f(x+n\eps)$. In fact, this program is only changing
the length of step. It is easy to see that all results in previous
sections will hold under the change of the length of step. We take
the limit $\eps\to 0$ then we get the classical AKNS-D hierarchy
which is briefly introduced in section II. The Taylor extension in
the proof of theorem 3 will give the standard relation between
$\partial_x$ and $\partial_{k\al}$ under this limit, i.e.
$\partial_x=\sum^m_{\al=1}a_{\al}\partial_{1\al}$
\cite{dic1,dic2,dic3}.

\section*{Acknowledgement}
This work is supported by K.C.Wong Education Foundation, Hong
Kong. The author would like to thank Prof. K.Wu for his helpful
discussion.



\begin{thebibliography}{99}
\bibitem{BS}A.~I.~Bobenko and R.~Seiler, {\it Discrete Integrable
Geometry and Physics}, Oxford University Press, 1999.

\bibitem{WZZ}Z.~Y.~Wu, D.~H.~Zhang and Q.~R.~Zheng, J. Phys. A :Math. Gen. {\bf 26}(1993) 2389.

\bibitem{MS}J.~Mac and M.~Seco, J. Math. Phys. {\bf 37}(1996) 6510.

\bibitem{FR}E.~Frenkel and N.~Reshetkhin, Commun. Math.
Phys.{\bf 178}(1996) 237.

\bibitem{KLR}B.~Khesin, V.~Lyubashenko and C.~Roger, J. Func.
Anal.{\bf 143}(1997) 55.

\bibitem{HI}L.~Haine and P.~Iliev, J. Phys. A : Math. Gen. {\bf
30}(1997) 7217.

\bibitem{Iliev1}P.~Iliev, Lett. Math. Phys. {\bf 44} No.3
(1998) 187.

\bibitem{TCL}M.~H.~Tu, J.~C.~Shaw and C.~R.~Lee, Lett. Math. Phys. {\bf
49}(1999) 33.

\bibitem{WWWY}S.~K.~Wang, K.~Wu, X.~N.~Wu and D.~L.~Yu, J. Phys. A : Math. Gen. {\bf 34} (2001) 9641.

\bibitem{Iliev2}L.~Haine and P.~Iliev, IMRN, No.6 (2000) 281.

\bibitem{dic1}L.~A.~Dickey, J. Math. Phys. {\bf 32} (1991) 2996.

\bibitem{dic2}L.~A.~Dickey, {\it Soliton Euqations and Hamiltonian
System}, Advenced Series in Mathmetical Physics, Vol.{\bf 12},
World Scientific Press, 1991.

\bibitem{dic3}L.~A.~Dickey, On Segal-Wilson's definition of
$\tau$-function of hierarchy AKNS-D and mcKP, in {\it Integrable
Systems : The Verdier Memorial Conference}, Ed. by O.~Babelon,
P.~Cartier and Yvette Kosmann-Schwarzbach, Birkhaeuser, Boston,
1993.

\end{thebibliography}
\end{document}